\documentclass[prb,preprint,superscriptaddress,amsmath,amssymb]{revtex4}
\usepackage{natbib}
\usepackage{graphicx}         
\usepackage{dcolumn}         
\usepackage{bm}                 

\begin{document}

\title{On the Possibility of Using Semiconductor Nanocolumns for the Realization of Quantum Bits}

\author{K.~M.~Indlekofer}
\email{m.indlekofer@fz-juelich.de}
\author{Th.~Sch\"apers}
\affiliation{Institute of Bio- and Nanosystems (IBN-1) and Center of
Nanoelectronic Systems for Information Technology (CNI), Research
Centre J\"ulich GmbH, D-52425 J\"ulich, Germany}

\date{\today}

\begin{abstract}
We propose the use of quantum dots formed in a semiconductor
nanocolumn for the realization of charge or spin based quantum bits.
The radial carrier confinement is achieved by employing conformal
overgrowth, while multiple segmented gates are used to control the
quantum dot properties. Different concepts for read-out and control
are discussed. Furthermore, we assess which combinations of core
nanowires and shell materials are feasible.
\end{abstract}

\maketitle

\section{Introduction}

Among the numerous concepts to implement a quantum circuit in a
quantum mechanical system, approaches based on semiconductor
quantum dots offer the great advantage that ultimately a
miniaturized version of a quantum computer is feasible. Various
approaches are pursued to realize a quantum dot quantum bit
(qubit). One possibility is to use the charge state to constitute
a qubit. Using this scheme coherent manipulation of a charge qubit
has been reported. \cite{Hayashi03,Gorman05,Fujisawa06} However,
regarding this approach it is not yet clear if the decoherence
times are long enough to allow a sufficiently large number of
quantum gate operations. Kane proposed to use the nuclear spin
state of a single $^{31}$P donor in a isotropically pure silicon
$^{28}$Si matrix to define a qubit.\cite{Kane98a} Since then, a
number of other approaches using Si:P, which rely on the charge
state \cite{Hollenberg04} as well as on the spin state
\cite{Vrijen00,Hill05} had been proposed. Experimentally,
considerable progress had been made to implement a Si:P-based
quantum computer.\cite{Obrien01,Buehler05,Buehler06,Stegner06}

A quantum computer based on the spin state in a semiconductor
quantum dots was first proposed by Loss and
DiVincenzo.\cite{Loss98} Here, the qubit is represented by a
single electron in a quantum dot. A semiconductor quantum dot can
be realized by using a two-dimensional electron gas (2DEG) and
split-gate electrodes defined by electron beam lithography. The
two-level system required for the qubit is obtained by inducing a
Zeeman spin splitting by applying a sufficiently large magnetic
field. Quantum mechanical superposition states can be achieved by
means of external electro-magnetic pulses, similar to electron
spin resonance transitions. Coupling between quantum dots, in
order to implement a two-qubit gate operation, can be realized by
controlling the gates, which separate two adjacent qubits.

An important prerequisite for the implementation of a qubit
following the approach of Loss and DiVincenzo is, that one is able
to trap a single electron in a quantum dot. This has been achieved
for planar quantum dots based on split-gates\cite{Ciorga00} as well
as for vertical quantum dots defined in resonant tunneling
structures.\cite{Tarucha96,Griebel98,Foerster98,Griebel99,Indlekofer00,Indlekofer02}
Recently, considerable progress has been made in demonstrating
physical effects that are important for the implementation of a
quantum circuit using this concept:\cite{Engel05,Cerletti05,Coish06}
i.e. the determination of spin relaxation time in a quantum dot,
\cite{Fujisawa02,Hanson03,Petta05a} the read-out of a single spin in
a quantum dot,\cite{Elzerman04,Hanson05} and the real-time detection
of a single electron by a quantum point contact.\cite{Vandersypen04}
Very recently, coherent oscillations of a single spin were
reported.\cite{Koppens06} Regarding the realization of two-qubit
quantum gates, coupled quantum dots have been investigated. Here,
coherent manipulation of coupled electron spins\cite{Petta05} as
well as control of tunnel splitting\cite{Huettel05} have been
achieved.

In recent years many novel methods have been developed to prepare
one-dimensional semiconductor structures. Beside planar wires
defined by etching or gating of semiconductor heterostructures, it
is also possible to prepare vertical one-dimensional structures
(nano\-columns) by etching. This so-called \emph{top-down} approach
was successfully used to prepare nanocolumn resonant tunneling
devices or vertical quantum
dots.\cite{Reed88,Kouwenhoven01,Indlekofer02,Wensorra05} In
addition, it is also possible to prepare vertical columns directly
by vapor-liquid-solid (VLS) epitaxy. This approach, often named
\emph{bottom-up} approach, was first established by Wagner
\cite{Wagner60} and was later refined to columns with diameters in
the nanometer range by Hiruma and coworkers.\cite{Yazawa91} Using
this method nanocolumns from different III-V semiconductors have
been prepared by chemical beam epitaxy (CBE) or by metal-organic
vapor phase epitaxy
(MOVPE).\cite{Samuelson04,Seifert04,Khorenko04,Roest06} The
functionality of the nanocolumns was enhanced considerably by
exploiting the growth of heterstructures in axial and radial
directions.\cite{Bjork02,Bjork02a,Gudiksen02,Seifert04,Verheijen06,Minot07}
Using VLS grown nanocolumns, the realization of various device
structures, i.e. resonant tunneling diodes, transistors or quantum
dots has been
demonstrated.\cite{Cui01,Lind06,Bryllert06,Do06,DeFranceschi03,Fasth05}

\section{Nanocolumn qubit system}

We propose the experimental realization of a qubit system via a
one-dimensional (1D) semiconductor heterostructure with conformal
overgrowth in a planar geometrical configuration with multiple
segmented gates for real-time control. The suggested system design
is suitable for both, charge and spin based qubit implementations.
For the latter option, the qubit is implemented either by the spin
of a single electron,\cite{Loss98,Elzerman04,Hanson05} or by the
singlet/triplet states of a coupled two-electron
system.\cite{Petta05a} The real-time control and coupling/separation
of individual quantum dot qubits is obtained with the help of
variable tunnel barriers, realized by gates and optionally built-in
heterostructures. A well-defined electronic filling (preparation) of
the system is accomplished via modulation doping and electronic
injection (single-electron tunneling) from contacts at the outer
channel terminals and with the help of suitable bias offsets at the
gate electrodes. Sequences of time-dependent voltage pulses at the
gate electrodes and outer contacts are used to manipulate the
electronic state of the system, optionally in combination with
packets of microwave excitation pulses and inelastic light
scattering (Raman). Furthermore, external magnetic fields and the
Rashba effect\cite{Bychkov84,Nitta97,Engels97} can be used to
control the electron spin. In this concept, the outer contacts are
also employed as probes for the measurement of electronic
occupation.
\begin{figure}
\resizebox{0.5\columnwidth}{!}{
\includegraphics{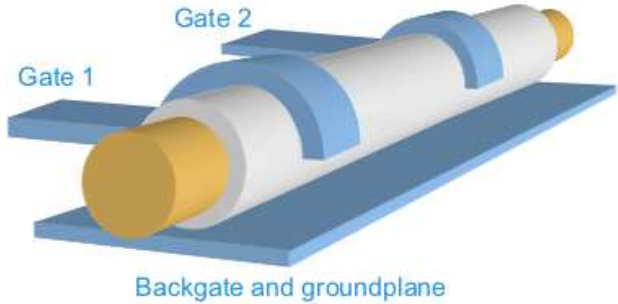}}
\caption{Schematic illustration of a planar quantum dot based on
conformally overgrown nanocolumns. Segmented semi-coaxial gates are
used to define and control the quantum dot.} \label{fig:1}
\end{figure}

\begin{figure}
\resizebox{0.5\columnwidth}{!}{
\includegraphics{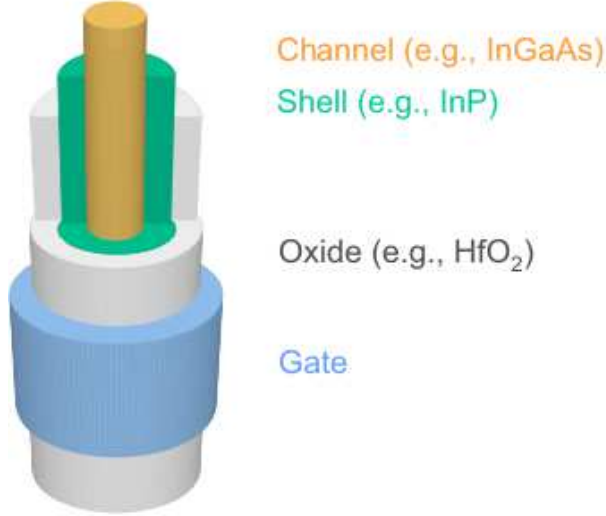}}
\caption{Layout of nanocolumn heterostructure
field-effect-transistor.  As an option, the a coaxial gate structure
is insulated from the semiconductor by an oxide layer.}
\label{fig:2}
\end{figure}

The system layout, as depicted in Fig.~\ref{fig:1}, resembles a
planar 1D field-effect-transistor (FET) structure with multiple
segmented Schottky gates (metal-semiconductor
field-effect-transistor: MESFET). As for the geometrical
configuration of the gate electrodes, a coaxial or semi-coaxial
arrangement will be employed, supplemented by a common back-gate
electrode. In the envisioned final realization, the transistor
channel is implemented as a semiconductor nanocolumn
heterostructure.\cite{Seifert04} Here, the nanocolumn may contain
longitudinal heterostructures for the pre-definition of quantum
wells and barriers. Two basic system types exist: A nominally
undoped nanocolumn with a conformal overgrowth of a barrier material
(e.g., AlGaAs on GaAs, or InP on InGaAs) with modulation doping can
be used, providing a coaxial 1D modulation-doped
field-effect-transistor (MODFET) with Schottky gates. Alternatively,
a 1D metal-oxide-semiconductor field-effect-transistor (MOSFET)
design might also be considered, either in a columnar or FinFET
\cite{Hisamoto00} geometry. Compared to the MESFET and MOSFET
design, the MODFET has the advantage that smaller inner channel
regions can be realized. Here, the channel region consists of a
radial quantum well which is defined via the core/shell
heterostructure leading to a strong and uniform radial confinement.
A further advantage is the perfect epitaxial heterointerface between
channel and barrier material. Combining the advantages of both
approaches, the metal-oxide-semiconductor MODFET design (also
referred to as a MOSHFET \cite{Marso06}), as depicted in
Fig.~\ref{fig:2}, employs an additional oxide insulation layer
between the semiconducting barrier material and the gate electrode
and thus provides an improved gate leakage behavior similar to a
MOSFET layout.

The proposed system design combines the advantages of planar
technology with those of vertical, epitaxially defined nanocolumn
structures. Here, the nanocolumn as a self-organized
\emph{bottom-up} structure exhibits an almost perfect crystalline
quality compared to etched bulk material (\emph{top-down}). In
addition, the use of epitaxial methods allows for tailored
electronic properties of the system in terms of longitudinal
heterostructures and doping profiles as a means to pre-define
quantum wells and barriers, in contrast to carbon-nanotubes and
2DEG-based systems. One important advantage of \emph{bottom-up}
nancolumns consists in the possibility to combine various, even
mismatched, materials. Furthermore, in contrast to 2DEG-based
systems, the radial confinement within the 1D structure is defined
inherently without the need for extra gates, which effectively saves
one dimension of confinement. Hence, the realization of one- or
two-electron states, which are typically required for a single qubit
system, becomes much easier with such a 1D
system.\cite{Indlekofer05,Indlekofer06} In addition, a coplanar
contact layout can be used for better high-frequency properties, as
compared to stacked gates in a vertical design with parasitic
capacitances. The planar gate layout also offers a better
scalability and provides a simpler processing technology for a
multi-dot and multi-gate architecture compared to a vertical design
of stacked gate layers, which require multiple processing steps.
Finally, an improved electrostatic control of the channel can be
achieved due to the coaxial or semi-coaxial gate geometry. The
proposed approach fulfills all basic criteria for a qubit
realization, since large parts of the concept are analogous to
existing 2DEG-based qubit concepts.

\section{Read-out, Calibration, and Initialization}

\begin{figure}
\resizebox{0.55\columnwidth}{!}{
\includegraphics{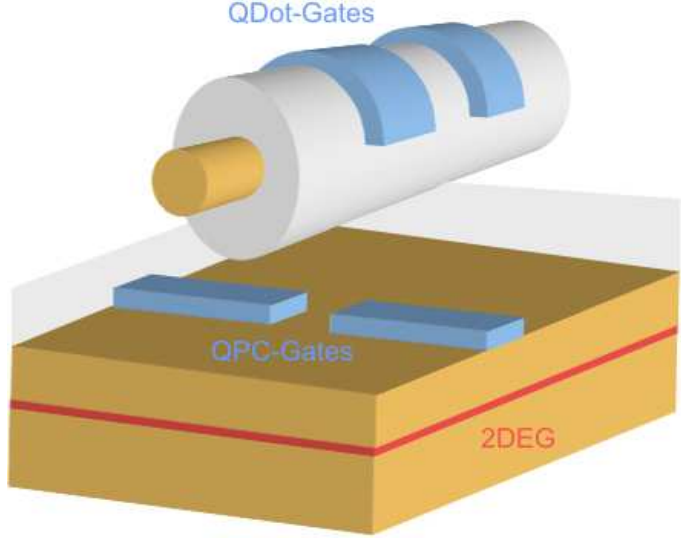}}
\caption{Read-out of a nanocolumn quantum dot qubit by a split-gate
quantum point contact underneath the quantum dot. Analogously, a
single-electron-transistor can be used as well.} \label{fig:3}
\end{figure}

\begin{figure}
\resizebox{0.5\columnwidth}{!}{
\includegraphics{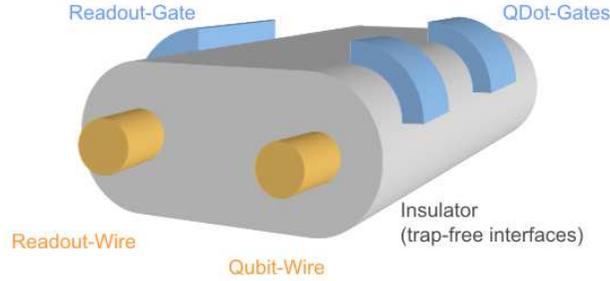}}
\caption{Alternative read-out configuration with an interacting
adjacent nanocolumn with a gate-defined constriction. In this
configuration, coalescence of the conformal shell material is
employed.} \label{fig:4}
\end{figure}

\begin{figure}
\resizebox{0.75\columnwidth}{!}{
\includegraphics{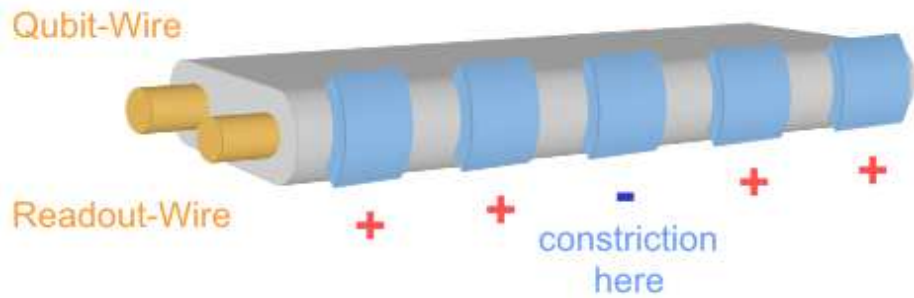}}
\caption{Multi-gate configuration of the dual-column system.}
\label{fig:5}
\end{figure}

Various approaches for the read-out, calibration, and initialization
processes can be pursued. As a first option, directly involving the
outer contacts, charge transport can be used as a means to detect
the occupation of single-electron levels or to probe singlet/triplet
states via spin-blockade. Alternatively, a spatially resolved charge
detection can be achieved by use of nanoscale electrometers based on
single-electron transistors (SET) or quantum point contacts (QPC).
As for the latter case, Fig.~\ref{fig:3} illustrates an embedded
QPC, realized in a 2DEG structure underneath the nanocolumn.
Alternatively a 1D FET/single-electron-transistor within a second
parallel nanocolumn can be employed, as depicted in
Fig.~\ref{fig:4}.
Here, multi-segment gate electrodes allow for the calibration of the
single-electron-transistor operation and the variable spatial
definition of sensitivity (cf. Fig.~\ref{fig:5}).
The second variant, the dual-nanocolumn arrangement, can be realized
via coalescence of the radial heterostructure during epitaxial
growth. As a major advantage, the usage of epitaxial interfaces
leads to the elimination of uncontrollable traps or interface
states. Static charges, which can be accounted for during
calibration of gate voltages, are unproblematic, in contrast to
non-reproducible states with adaptive charging (and hence
unintentional screening). As an option, additional intermediate
screening gates can be employed to provide spatially well-defined
excitation regions within the channel of the nanocolumn, which
becomes important with respect to the usage of high frequency gate
voltages (CW and pulsed).

\section{Material System and Dimensions}

The fabrication of the 1D semiconductor heterostructure and the
optional subsequent conformal overgrowth is accomplished in a
\emph{bottom-up} approach by use of molecular-beam epitaxy (MBE) or
metal-organic-vapor-phase epitaxy (MOVPE), and possible combinations
thereof. Here, channel doping profiles and modulation doping is
introduced during the growth process itself. Various semiconductor
material systems are suitable for the realization of a 1D qubit
system. One of the most promising material system is GaAs/AlGaAs,
where GaAs is used as the channel material and Al$_x$Ga$_{1-x}$As as
a barrier material between quantum dots within the channel as well
as for the conformal overgrowth. As a passivation for Al-rich
compounds, GaAs can be employed as an outer cap-layer.
Alternatively, InAs or Ga$_x$In$_{1-x}$As can be used as the channel
material with Al$_y$In$_{1-y}$As or InP as barrier material. The
InN/AlN/GaN material system has to be considered as an option.
\cite{Calarco05} Here, InN might be used as the channel
material.\cite{Chang05,Stoica06} However, the current crystalline
quality of nitride semiconductors might be not sufficient for
scalable quantum structures. Furthermore, the possibility of a
conformal growth mechanism for nitride semiconductors still has to
be investigated. Optionally, the Si/SiGe material system might be
considered as well, offering the most advanced device technology. In
this context, SiO$_2$ and alternative \emph{high-k} dielectrics
(e.g. HfO$_2$) are the first choice for gate insulators and
barriers. As compared to III/V technology, however, SiGe
heterostructures lack the flexibility in material combination. Here,
reasonable barrier heights for quantum well structures typically
require an extremely large lattice mismatch.

In addition to the epitaxially grown nanocolumns, nanotubes formed
by self-scolling might also be of interest.\cite{Prinz00,Zhang04}
This concept offers the big advantage to yield 1D nanostructures in
a planar configuration with pre-defined size, position, and
alignment.

Indium-based systems become advantageous for a large Zeeman and
Rashba splitting for spin-based qubits, employing the spin of a
single electron or singlet/triplet states of a two-electron system.
Furthermore, they exhibit a small effective mass which provides
increased quantization energies and, in turn, allows for larger
feature sizes. However, with the usage of In-based semiconductor
compounds one has to account for the reduced band gap and the
possible existence of surface-charge accumulation layers.

The best performance and scalability can be expected from the
conformally overgrown (MODFET-like) structures without doping within
the inner channel region, in order to suppress the direct influence
of individual impurities (and the resulting Anderson localization)
and furthermore to guarantee dopant ionization at low temperatures.
Nevertheless, tailored doping profiles can be considered as an
option to implement/pre-define quantum dots and barriers, e.g.
saddle-points in the three-dimensional potential.\cite{Wensorra05}
The role of alloy scattering on a nanoscale in ternary and
quaternary compound semiconductors has to be analyzed in this
context as well.

\section{Placement and Lithography}

In case of vertically grown nanocolumn structures, a crucial step is
the controlled transfer into a planar position. In a first approach,
one can employ suspended detached nanocolumns, spin-coat the target
substrate, and make use of a scanning electron microscope to locate
suitable candidates among the remaining randomly distributed
nanocolumns. With the help of this spatial information, the
subsequent electron beam lithography has to be adjusted to the
actual nanocolumn position and alignment; in particular, for the
definition of markers for following lithographic steps. Future
implementations might even allow for a pre-defined growth and
placement of the 1D nanocolumn structures. For example, one might
combine the fabrication steps for rolled-up structures with the
self-organized growth of nanocolumns. A combination of electron-beam
lithography and standard optical lithography can be used for the
definition of gate electrodes and the outer source and drain
contacts. Optionally, a conductive substrate or a patterned
metallization layer can be used as a common backgate beneath the
whole 1D nanostructure, serving also as a groundplane for the
high-frequency microstrip design. In order to separate the
nanocolumn and the electrodes from the substrate, insulating
dielectric layers have to be employed, patterned by lithographical
means. Source and drain contacts are either Schottky or ohmic
contacts. They are used for the preparation of the quantum state via
electronic injection from the source and drain contact regions
(reservoirs) and can be employed for the measurement of electronic
occupation. The source and drain contacts must be electronically
separable from the channel region, which is accomplished by means of
gates at the outer ends of the channel. For ohmic contacts on GaAs,
one might also consider non-alloyed contacts based on
low-temperature-grown-GaAs.\cite{Wensorra05,Wensorra06}

\section{Summary}

In summary, a quantum bit system based on quantum dots in
semiconductor nanocolumns has been proposed. The nanocolumns can be
prepared directly by epitaxial growth, e.g. by using vapor-liquid
solid epitaxy. Carrier confinement in radial direction can be
achieved by conformal overgrowth of a semiconductor barrier layer or
by covering the nanocolumn by a dielectric layer. In longitudinal
direction, the electrons are controlled by means of multiple
semi-coaxial or coaxial gate electrodes. The read-out process can be
implemented via a QPC or SET, realized within a 2DEG underneath the
nanocolumn or within an adjacent second nanocolumn. As for the
choice of a suitable material system, III/V semiconductors such as
AlGaAs/GaAs or GaInAs/InP are the most promising candidates, while
Si/SiGe in combination with a gate-dielectric (e.g. SiO$_2$ or
HfO$_2$) is also an option.


\end{document}